\documentclass[aps,prl,twocolumn,superscriptaddress]{revtex4-1}

\usepackage{amsmath}
\usepackage{graphicx}
\usepackage{dcolumn}
\usepackage{bm}
\usepackage{amssymb}
\usepackage[detect-all,separate-uncertainty = false,multi-part-units=single,list-units = single,range-units = single]{siunitx}
\usepackage{physics} 
\usepackage{color} 
\graphicspath{{figures/}}

\begin{document}


\title{New bounds from positronium decays on massless mirror dark photons}
\author{C.~Vigo}
\affiliation{ETH Zurich, Institute for Particle Physics and Astrophysics, 8093 Zurich, Switzerland}
\author{L.~Gerchow}
\affiliation{ETH Zurich, Institute for Particle Physics and Astrophysics, 8093 Zurich, Switzerland}
\author{B.~Radics}
\affiliation{ETH Zurich, Institute for Particle Physics and Astrophysics, 8093 Zurich, Switzerland}
\author{M.~Raaijmakers}
\affiliation{ETH Zurich, Institute for Particle Physics and Astrophysics, 8093 Zurich, Switzerland}
\author{A.~Rubbia}
\affiliation{ETH Zurich, Institute for Particle Physics and Astrophysics, 8093 Zurich, Switzerland}
\author{P.~Crivelli}
\affiliation{ETH Zurich, Institute for Particle Physics and Astrophysics, 8093 Zurich, Switzerland}
\email[]{crivelli@phys.ethz.ch}

\date{\today}

\begin{abstract}
We present the results of a search for a hidden mirror sector in positronium decays with a sensitivity comparable with the bounds set by the prediction of the primordial He$^{\num{4}}$ abundance from Big Bang Nucleosynthesis. No excess of events compatible with decays into the dark sector is observed resulting in an upper limit for the branching ratio of this process of \num{3.0E-5} (\SI{90}{\percent} C.L.). This is an order of magnitude more stringent than the current existing laboratory bounds and it constraints the mixing strength of ordinary photons to dark mirror photons at a level of $\varepsilon<\num{5.0E-8}$. 
\end{abstract}

\maketitle

\section{Introduction}
\label{Sec:intro}
``Now, after the first shock is over, I begin to collect myself. Yes, it was very dramatic.'' This extract of a letter  from W.~Pauli to V.~Weisskopf~\cite{Paulibiography} reflects the state of mind of the physical community in 1957 after the announcement of the discovery of parity violation in weak interaction by Wu~\cite{Wu:1957my} and Lederman~\cite{Garwin:1957hc} predicted by Lee and Yang~\cite{Lee:1956qn} one year before. Nowadays, parity violation is put "ad hoc" in the Lagrangian of the Standard Model (SM) in the framework of the vector $-$ axial vector (V$-$A) theory~\cite{PhysRev.109.193,PhysRev.109.1860.2}, but still there is no explanation of why the vacuum appears left-right asymmetric. 

Some models postulate the suppression of the reverse chirality (vector + axial vector, V+A) component in weak interaction by a heavy W$_\text{R}$ boson such that parity would be restored at high energies~\cite{Mohapatra:1974gc,Mohapatra:1974hk}. An alternative solution is the one already discussed by Lee and Yang in their original paper: in order to save parity conservation, the transformation in the particle space corresponding to the space inversion  ${\bf x\to -x}$ should not be the usual transformation  {\it  P} but  {\it  PR}, where  {\it  R} corresponds to the transformation of a particle into a reflected state in the mirror particle space.  

The idea that for each ordinary particle, such as the photon, electron, proton and neutron, there is a corresponding mirror particle of exactly the same mass and properties as the ordinary particle, was further developed over the years~\cite{Salam:1957st,Kobzarev:1966qya,Foot:1991bp,Berezhiani:1995yi,Blinnikov:1982eh}. R-parity interchanges the ordinary particles with the mirror particles. Parity is conserved because the mirror particles experience V$+$A (i.e.~right-handed) mirror weak interactions while the ordinary particles experience the usual V$-$A (i.e.~left-handed) weak interactions.  

Doubling the content of the Standard Model to solve some problems might seem un-natural, however it has worked in the past. From the union of quantum mechanics and relativity, anti-matter was postulated. 

Moreover, mirror matter being stable and massive is an excellent candidate for Dark Matter (DM). In fact, even though the existence of DM has been established by different cosmological observations (see e.g.~\cite{RevModPhys.90.045002} for a recent review), its origin is still unknown. Many candidates have been proposed among which the most popular one is the Weakly Interacting Massive Particles (WIMPs). Despite intensive searches in accelerators and in direct detection experiments~\cite{Arcadi:2017kky}, WIMPs have not yet been observed.  An interesting alternative, which gained a lot of attention in recent years, is hidden sectors~\cite{Alexander:2016aln,Battaglieri:2017aum}. This class of models  includes the possibility of a new force mediated by a massive vector gauge $U(1)$ boson, known as Dark Photon (A'). The A' would mediate the interaction from ordinary and hidden sectors via kinetic mixing $\mathcal{L} = \varepsilon F^{\mu\nu}F'_{\mu\nu}$ where $\varepsilon$ is the strength and $F^{\mu \nu}$ ($F'_{\mu \nu}$) the SM (hidden) electromagnetic field strength tensor. This term is gauge invariant and renormalizable. If the new $U(1)$ gauge symmetry is unbroken, the A' is massless and for the mirror hidden sector, the A' corresponds to the so called mirror photon~\cite{Holdom:1985ag}.

The photon mirror-photon kinetic mixing would break the degeneracy between the triplet spin state of the electron-positron bound state called orthopositronium (o-Ps) and its mirror partner (o-Ps')~\cite{Glashow:1985ud}, connected via the o-Ps virtual annihilation channel. The vacuum energy eigenstates are a linear combination of the mass eigenstates $(\text{o-Ps}\pm\text{o-Ps'} )/\sqrt{2}$, which are separated by an energy $\Delta = 2h\varepsilon \nu$, with $h$ the Planck constant and $\nu=\SI{8.7e4}{\mega\hertz}$ the contribution from the virtual one-photon orthopositronium decay channel~\cite{Berko1981787}. This would lead to orthopositronium to mirror orthopositronium Rabi oscillations. The probability of o-Ps being in its mirror matter state after a time $t$ is given by 
\begin{equation}
P(\text{o-Ps}\rightarrow\text{o-Ps'}) = \exp(-\Gamma_{\text{SM}} t)\sin^2\Omega t,
\end{equation}
with $\Omega = 2\pi \varepsilon \nu$ and $\Gamma_{\text{SM}}=\SI{7.040}{\per\micro\second}$ the o-Ps decay rate predicted by the SM~\cite{theoretical_ops} that has been  accurately measured~\cite{rich_ops, Kataoka2009}. 
The branching ratio for $\text{o-Ps}\rightarrow\text{o-Ps'}$ in vacuum is thus given by: 
\begin{eqnarray}
\label{eq:branching_ratio}
{\mathcal{BR} (\text{o-Ps}\rightarrow \text{o-Ps'})}  = &\Gamma_{\text{SM}}\int^\infty_0 P(\text{o-Ps'})\; dt \nonumber \\ 
= &\frac{2\Omega^2}{\Gamma_{\text{SM}}^2 + 4 \Omega^2}
\end{eqnarray}

The experimental signature of this process is the apparently invisible decay of o-Ps, such that the energy $2m_\text{e}$ expected for ordinary decays is missing in a hermetic calorimeter surrounding the o-Ps formation target. Therefore, the occurrence of the $\text{o-Ps}\rightarrow\text{o-Ps'}$ conversion would appear as an excess of events with zero-energy deposition in the calorimeter above the expected background.

Previous experiments searching for $\text{o-Ps}\rightarrow\text{invisible}$ have been performed with o-Ps confined in the pores of aerogels~\cite{Atoyan1989, Mitsui1993}. The most stringent limit on the branching ratio of this process is \num{4.2E-7}~\cite{Badertscher2007}. However, since collisions with matter destroy the coherence of the oscillation, the branching ratio of the $\text{o-Ps}\rightarrow\text{o-Ps'}\rightarrow\text{invisible}$ process is suppressed scaling approximately as the square root of the number of collision ($N_\text{coll}$). In the aerogel pores, o-Ps undergoes approximately $N_\text{coll}=\num{E4}$ collisions per lifetime, thus the oscillation is inhibited by a factor  $\sqrt{N_\text{coll}}\simeq\num{100}$. Therefore, an experiment in vacuum is much more sensitive to $\varepsilon$ and it allows to remove the systematic uncertainty related to o-Ps collisions in the pores as it was recently demonstrated~\cite{Vigo2018}.

An upper limit of $\varepsilon<\num{3E-8}$ was deduced by the successful prediction of the primordial He$^{\num{4}}$ abundance via Big Bang Nucleosynthesis~\cite{Carlson:1987si}. Considering additional cosmological and astrophysical observations such as cosmic microwave background and large scale structure formation, a more stringent bound at a level of \num{E-9} can be obtained~\cite{Berezhiani:2008gi,Foot:2014mia}. These values are in the range of naturally small $\varepsilon$ motivated by grand unification models~\cite{Berezhiani:2000gw} and by cosmology~\cite{Foot:2015}.

\section{Experimental method and setup}
\label{sec:setup}

\begin{figure}
	\centering
	\includegraphics[width=\linewidth]{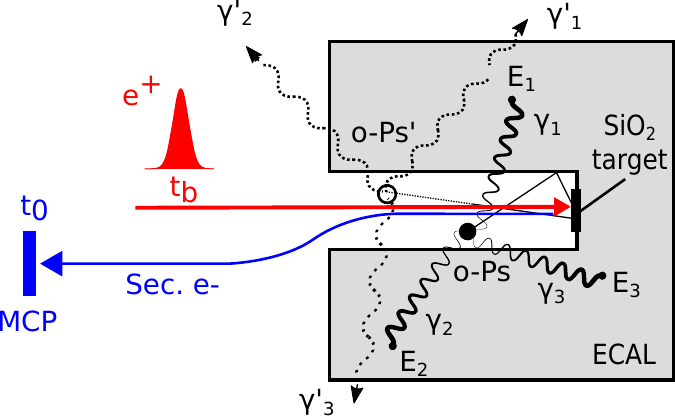}
	\caption{Sketch of the experimental setup and technique. ECAL granularity and vacuum pipe have been omitted for better visualization. See text for further details.}
	\label{fig:expsketch}
\end{figure}

\begin{figure}
	\centering
	\includegraphics[width=\linewidth]{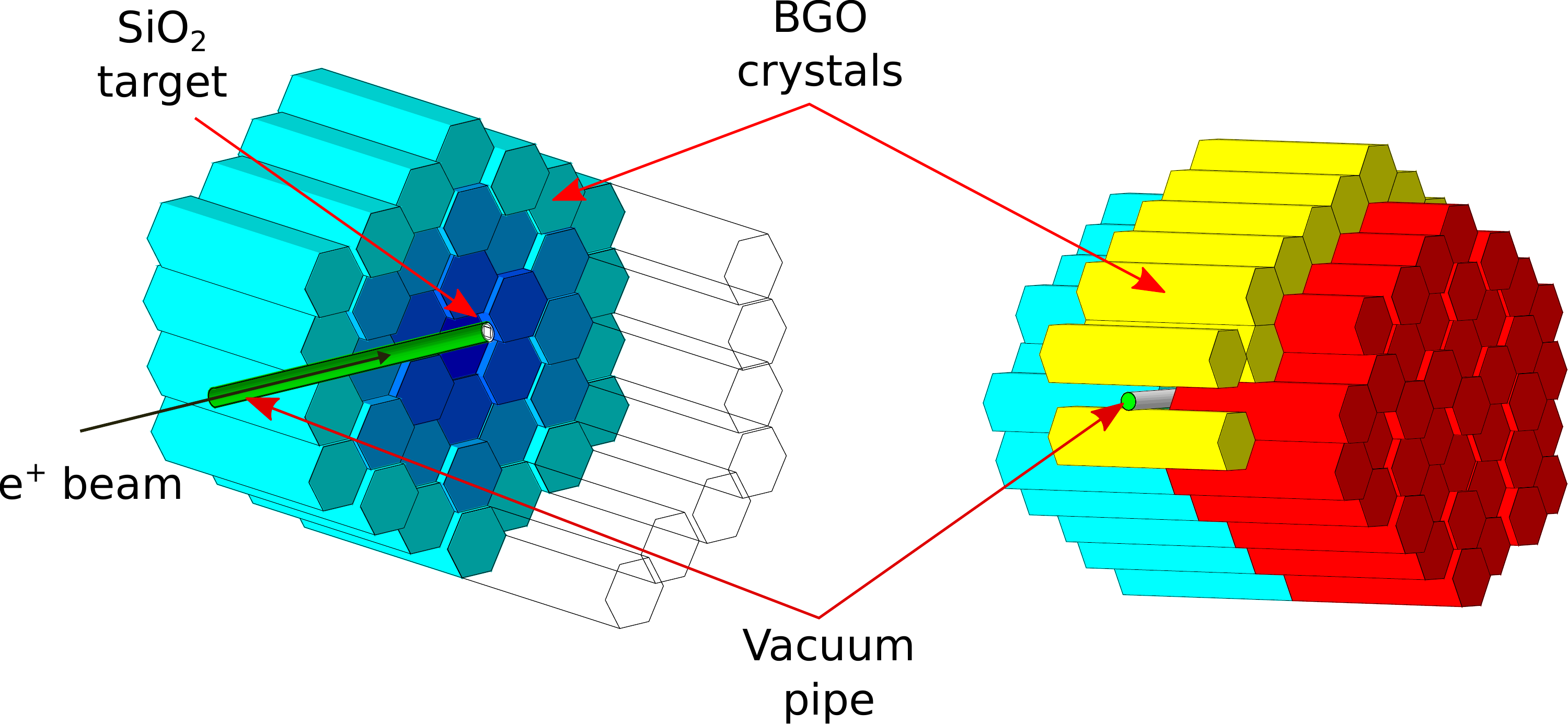}
	\caption{The ECAL surrounding the vacuum pipe and the porous silica target consists of \num{91} BGO crystals in a honeycomb structure providing granularity and high hermeticity.}
	\label{fig:BGO_enclosure}
\end{figure}

The principle of the experiment is sketched in Fig.~\ref{fig:expsketch}. Positrons from the ETH Zurich beam~\cite{Vigo2018} impinging on a porous silica target produce positronium emitted into vacuum with a conversion efficiency of about \SI{30}{\percent}~\cite{Liszkay2008}.
The target is surrounded by \num{91} bismuth germanium oxide (BGO) crystals forming a highly hermetic calorimeter (ECAL) as shown in Fig.~\ref{fig:BGO_enclosure}. When positrons hit the target secondary electrons (SE) are released. The SE are guided to a micro-channel plate (MCP), providing the tagging of the positron arrival at the target and opening a gate of \SI{3}{\micro\second} for the ECAL data acquisition. To decrease the number of accidental triggers the positron beam is chopped to \SI{300}{\nano\second} wide pulses at a rate of \SI{333}{\kilo\hertz} and then time-compressed (bunched) using a time dependent potential, resulting in a narrower distribution in the arrival of positrons at the target defined as the time difference between the chopper gate opening ($t_\text{b}$) and the arrival of a secondary electron ($t_0$).

The energy deposited by the annihilation photons is recorded for each individual BGO crystal. Some of the BGO signals are split to record also the time distributions. This allows us to determine the fraction of o-Ps emitted into vacuum and monitor it during the run by fitting the time spectra to extract the intensity of the long lived $\approx\SI{142}{\nano\second}$ component~\cite{Liszkay2008}.

An event is considered to be zero-energy compatible when each individual crystal $k$ measures an energy below a threshold $E_k^\text{T}$. The thresholds $E_k^\text{T}$ are determined with the incoming positron beam shut off to ensure a total detection efficiency $\eta_\text{total}>0.9$. For a more detailed description of the setup, the reader should refer to~\cite{Vigo2018,VigoPhD}. Here we report two major upgrades that increased the signal-to-background ratio by one order of magnitude. 

To form positronium emitted into vacuum with a kinetic energy of $\simeq\SI{100}{\milli\electronvolt}$, the positrons have to be implanted in the porous silica film with energies of few \si{\kilo\electronvolt}. This is done by biasing the target at high potential of typically \SIrange{1}{3}{\kilo\volt}. In the previous search~\cite{Vigo2018}, this resulted in a significant emission of electrons inducing false triggers in the system (uncorrelated with positrons) and thus limiting the sensitivity of the experiment.

\begin{figure}
	\centering
	\includegraphics[width=\linewidth]{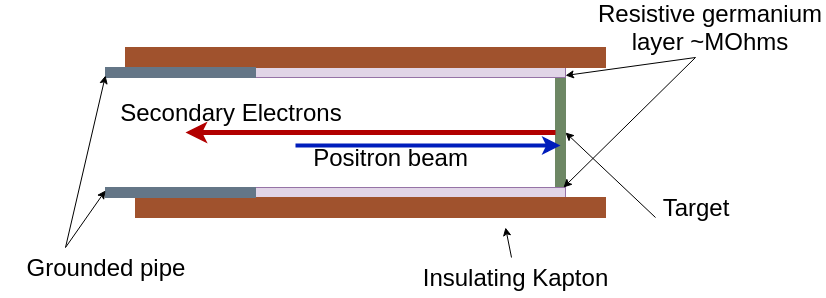}
	\caption{A sketch of the newly designed cavity. The high resistance germanium layer connecting the target with the grounded pipe reduces the extreme fields near the target. }
	\label{fig:cavity_design}
\end{figure}

The target design has been improved to reduce the effects of large electric fields near the target due to the high potentials applied to it.  A sketch of the new design is shown in Fig.~\ref{fig:cavity_design}.

In the previous cavity design the target was separated from the grounded pipe only by a thin layer of Kapton, resulting in strong electric fields ($\approx\SI{4}{\kilo\volt\per\centi\metre}$). In the new design a thin \SI{40}{\nano\metre} germanium layer of \SI{40}{\milli\metre} length was added to connect the target with the grounded pipe. In this way, the electric fields are reduced to $\approx\SI{0.5}{\kilo\volt\per\centi\metre}$ and the electron emission at \SI{3}{\kilo\volt} by a factor \num{100}.

The tungsten moderator has been upgraded to an argon moderator, increasing the average positron flux by \num{20}. Unlike the Gaussian beam profile of the tungsten moderator, the new moderator produces a donut-shaped profile arising from the conical shape of the source holder used to optimize the positrons moderation efficiency. Therefore, the secondary electrons are released closer to the walls of the cavity, reducing the tagging efficiency by a factor of \num{2}.

Two types of background contribute to the number of zero-energy compatible events, namely uncorrelated and correlated with the positrons arrival at the target.
\begin{enumerate}
\item[i)] Uncorrelated background, or false triggers, are uniformly distributed in time and originate from dark counts in the MCP or electron emission from the target. Dark counts in the MCP are approximately \SI{0.1}{\hertz}. Electron emission is strongly dependent on the set potential. At \SI{2}{\kilo\volt} this is less than \SI{1}{\hertz}, while at a potential of \SI{3}{\kilo\volt} it increases to \SI{50}{\hertz}. The contribution from false triggers is measured from the data by using a control region away from the positron pulse arrival.

\item[ii)] Background correlated with the positron arrival at the target arises from backscattered positrons, o-Ps escaping the detection region and ECAL hermeticity. Backscattered positrons is the dominant contribution, the other two sources being expected below \num{E-6}~\cite{Vigo2018} and \num{E-7}~\cite{VigoPhD} respectively. Positrons impinging on the target backscatter with a probability ranging from \SIrange{1}{2}{\percent} at \SI{1}{\kilo\electronvolt}, decreasing with the implantation energy~\cite{Gidley1995}. If the positron loses less energy than the one it had before being accelerated by the target voltage, i.e.~the energy corresponding to the source bias and the additional energy gained during the bunching process, it may escape the detection region and result in a zero-energy event. A rejection electrode can be used to prevent this as demonstrated in~\cite{Vigo2018}, however in the measurements presented here this was not working due to technical issues. A Monte-Carlo simulation was developed to calculate the expected background from this process. This required a detailed description of the electric and magnetic fields computed with the SIMION software~\cite{SIMION} and then imported into the GEANT4 simulation~\cite{Geant4}. The resulting escape probability $P_\text{fast}$ is \numrange{1}{1.7E-5}. The main sources of systematic uncertainties are the positron energy distribution ($\pm\SI{50}{\electronvolt}$ due to the bunching process) and the shape of the beam profile. Assuming the two to be uncorrelated gives a total uncertainty of \SIrange{35}{40}{\percent}.
\end{enumerate}

\section{Data analysis and results}

\begin{figure*}
  \centering
  \includegraphics[width=1.\textwidth]{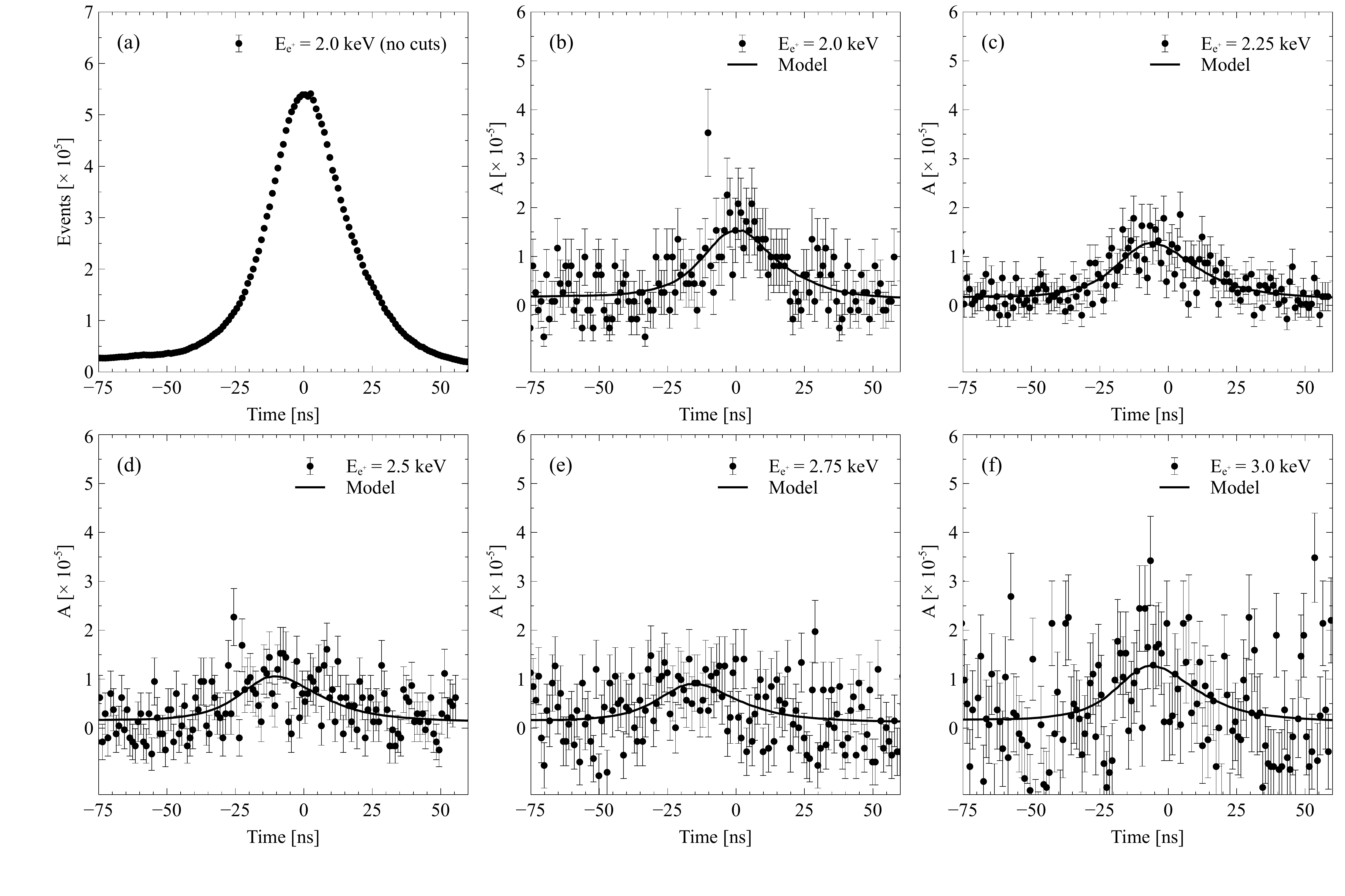}
  \caption{(a) Time distribution of the positron pulse at the target for $E_{\text{e}^+}=\SI{2}{\kilo\electronvolt}$. (b)-(f) Measured zero-energy pulse shape after subtraction of the flat background contribution arising from accidentals normalized to the total number of positrons on target.}
\label{fig:pulseshape}

\end{figure*}

The data were taken at $E_{\text{e}^+}=$ \SIlist{2.0;2.25;2.5;2.75;3}{\kilo\electronvolt} positron implantation energies. Fig.~\ref{fig:pulseshape}~(a) shows the time distribution of positron arrivals at the target for the \SI{2}{\kilo\electronvolt} data point (this is representative for all other energy points).  

Since any production of a signal or a signal-like background must be completely driven by the positron arrival at the target, we model the shape of both the irreducible background and the expected signal following the measured pulse shape, as shown in Fig.~\ref{fig:pulseshape}~(a).

The plots in Fig.~\ref{fig:pulseshape}~(b)-(f) show the time distribution of the fraction of zero-energy events, denoted as $A$, normalized to the total number of positrons on target after subtraction of the flat background contribution arising from accidentals. The zero-energy peak decreases with increasing implantation energy as expected for the background. For a signal arising from o-Ps' decays, one expects the opposite behaviour. In fact, increasing the implantation energy decreases the number of collisions with the vacuum cavity and thus enhances the oscillation probability.

The time distribution at \SI{3}{\kilo\electronvolt} implantation energy shows large statistical fluctuations due to a higher background from accidentals. This is expected because electron emission from the target increases greatly at larger electric fields, as mentioned in the previous Section. This data set was thus excluded from any further analysis.

\begin{figure}
\centering\includegraphics[width=1\linewidth]{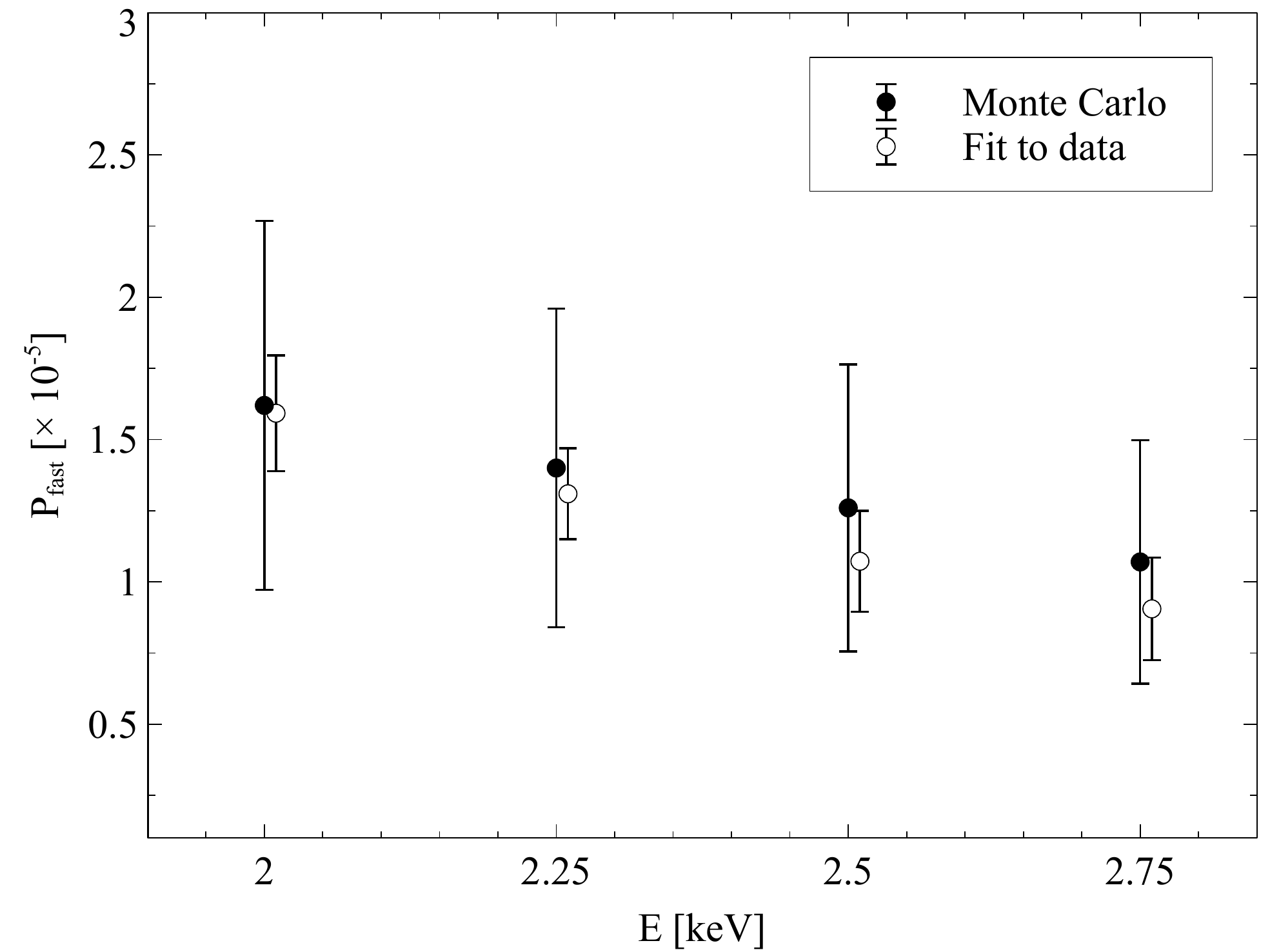}
\caption{Background estimation due to fast back-scattered positrons from MC simulations (full circles) and observed fraction of zero-energy compatible events (empty circles). The small offset in implantation energy between data points is artificially added for visualization purposes.}
\label{fig:data_bkgnd}
\end{figure}

 Fig.~\ref{fig:data_bkgnd} shows that the fractions of zero-energy events ($A$) extracted from Fig.~\ref{fig:pulseshape}~(b)-(e) are compatible with the escape probabilities $P_\text{fast}$ predicted by the Monte-Carlo simulation within the systematic uncertainties, plotted as error bars.

The analysis of the data is performed using a Bayesian approach. The four datasets are fit jointly by a likelihood model, with a statistical term
\begin{eqnarray}\label{eq:likelihood}
P\left(\vec{N}|B_\text{acc}, P_\text{fast},\mathcal{BR}_{\text{s}}\right)  & = {\displaystyle \prod_{i}^{\mathrm{bins}}} \frac{\left[B_\text{acc}+B_\text{fast}(i)+S(i)\right]^{n(i)}}{n(i)!} \nonumber \\ 
& \times e^{-(B_\text{acc}+B_\text{fast}(i)+S(i))} 
\end{eqnarray}
where $P$ is the joint probability of observing the data $\vec{N} = \left\lbrace n_i\right\rbrace$ given the following model:
\begin{enumerate}
\item[a)] $B_{\text{acc}}$ is the flat, uncorrelated background arising from accidentals.
\item[b)] $B_\text{fast}(i) = P_\text{fast}\cdot\eta\cdot \Phi(i)$ is the background due to fast back-scattered positrons in time bin $i$, with $P_\text{fast}$ the escape probability, $\Phi(i)$ the positron flux in time bin $i$, $n_i$ the number of observed events in time bin $i$ and $\eta=0.91\pm0.01$ the detector efficiency.
\item[c)] $S(i)=\mathcal{BR}_{\text{s}}\cdot K'\cdot \Phi(i)$ is the expected signal with branching ratio $\mathcal{BR}_{\text{s}}$, in time bin $i$, where $K' = \eta \cdot f_{\text{o-Ps}}\cdot \alpha_E$, with $f_{\text{o-Ps}}=0.30\pm0.02$ the fraction of o-Ps and $\alpha_E = \left\lbrace0.48, 0.51, 0.57, 0.59\right\rbrace$ the suppression factor for $E = \left\lbrace2, 2.25, 2.5,2.75\right\rbrace\si{\kilo\electronvolt}$ implantation energies. $\alpha_E$ takes into account the suppression of the oscillation of oPs$\to$oPs' arising from the decoherence induced by the collisions. This is calculated by numerically solving the differential equations governing the oscillation process including this effect~\cite{Vigo2018}. The measured positronium velocity distributions~\cite{Crivelli2010_2,Cassidy2010} are used as an input for MC simulations to estimate the mean number of o-Ps collisions with the vacuum walls as a function of $E$. The systematic uncertainty originating from the uncertainty in the measured energy and angular distributions is estimated to be less than \SI{10}{\percent} for all the energy data points.
\end{enumerate}

We use Bayes theorem to fit the data with the statistical likelihood term in Eq.~\ref{eq:likelihood}, Gaussian penalty terms for the systematic uncertainties and a flat prior on the signal branching ratio. The model parameters $P_\text{fast}$ and $B_\text{acc}$ are separately fit in each dataset, while $\mathcal{BR}_{\text{s}}$ is the single joint parameter used in the fit of all of the datasets.

No excess above the background is found, thus we set upper limits (at \SI{90}{\percent} C.L.) for the branching ratios of each of the following processes:
\begin{enumerate}
\item[a)] Positron invisible decay: $\mathcal{BR}\left(\text{e}^{+}\rightarrow\text{invisible}\right)<\num{5.0E-6}$
\item[b)] Model-independent o-Ps$\to$inv.~process (i.e.~$\alpha_E = \left\lbrace1,1,1,1\right\rbrace$): $\mathcal{BR}\left(\text{o-Ps}\rightarrow\text{invisible}\right)<\num{1.7E-5}$
\item[c)] $\mathcal{BR}\left(\text{o-Ps}\rightarrow\text{o-Ps'}\rightarrow \text{invisible}\right)<\num{3.0E-5}$
\end{enumerate}
By solving Eq.~\ref{eq:branching_ratio} as a function of the mixing strength, this last limit is converted in an upper bound of $\varepsilon<\num{5.0E-8}$.

For a background-only model ($\mathcal{BR}_\text{s} = 0$) fit to the data, we obtain from a goodness-of-fit test a value of $\chi^2/\text{NDF} = \num{1.08}$ ($\text{NDF} = \num{552}$).

\section{Conclusions}
No excess of zero-energy compatible events out of \num{2E8} tagged positrons is observed above the expected background extracted from the data, thus a ${\mathcal{BR}\left(\text{o-Ps}\rightarrow\text{o-Ps'}\rightarrow\text{invisible}\right)}<\num{3.0E-5}$ at \SI{90}{\percent} C.L.~is set. This result excludes the existence of massless mirror dark photons coupling through kinetic mixing with a strength larger than $\varepsilon<\num{5.0E-8}$.

This is the first time that a laboratory experiment reaches a sensitivity comparable with cosmological bounds. Future improvements of this experiment include a neon based moderator and a re-moderation stage in transmission mode~\cite{NiRem}. The overall trigger rate will be reduced by a factor 3 but the much better quality of the positron beam (i.e.~much smaller energy spread) will allow to improve the bunching compression and thus the signal-to-background ratio by at least a factor of 10. 
The implementation of a carbon foil will prevent o-Ps escaping the detection region and improve the confidence level of the tagging system, reducing the background from electron emission by two orders of magnitude~\cite{Crivelli2010}. This will allow to push the experiment to the ultimate sensitivity of $\varepsilon\approx\num{E-9}-\num{E-10}$, which is of great interest both theoretically and phenomenologically. 

\section*{Acknowledgments}
We express our deep gratitude to S.~Gninenko and A.~Belov for their essential contributions to the first stages of the experiment. We also wish to thank R.~Vallery and D.~Cooke for their valuable discussions and very helpful comments. This work was supported by the ETH Zurich Grant No.~ETH-35-14-2.

\bibliographystyle{elsarticle-num}
\bibliography{bibliography.bib}

\begin{thebibliography}{10}
\expandafter\ifx\csname url\endcsname\relax
  \def\url#1{\texttt{#1}}\fi
\expandafter\ifx\csname urlprefix\endcsname\relax\def\urlprefix{URL }\fi
\expandafter\ifx\csname href\endcsname\relax
  \def\href#1#2{#2} \def\path#1{#1}\fi

\bibitem{Paulibiography}
C.~P. Enz, No Time to be Brief, Oxford University Press, 2002.

\bibitem{Wu:1957my}
C.~S. Wu, E.~Ambler, R.~W. Hayward, D.~D. Hoppes, R.~P. Hudson, {Experimental
  Test of Parity Conservation in Beta Decay}, Phys. Rev. 105 (1957) 1413--1414.
\newblock \href {https://doi.org/10.1103/PhysRev.105.1413}
  {\path{doi:10.1103/PhysRev.105.1413}}.

\bibitem{Garwin:1957hc}
R.~L. Garwin, L.~M. Lederman, M.~Weinrich, {Observations of the Failure of
  Conservation of Parity and Charge Conjugation in Meson Decays: The Magnetic
  Moment of the Free Muon}, Phys. Rev. 105 (1957) 1415--1417.
\newblock \href {https://doi.org/10.1103/PhysRev.105.1415}
  {\path{doi:10.1103/PhysRev.105.1415}}.

\bibitem{Lee:1956qn}
T.~D. Lee, C.-N. Yang, {Question of Parity Conservation in Weak Interactions},
  Phys. Rev. 104 (1956) 254--258.
\newblock \href {https://doi.org/10.1103/PhysRev.104.254}
  {\path{doi:10.1103/PhysRev.104.254}}.

\bibitem{PhysRev.109.193}
R.~P. Feynman, M.~Gell-Mann,
  \href{https://link.aps.org/doi/10.1103/PhysRev.109.193}{Theory of the fermi
  interaction}, Phys. Rev. 109 (1958) 193--198.
\newblock \href {https://doi.org/10.1103/PhysRev.109.193}
  {\path{doi:10.1103/PhysRev.109.193}}.
\newline\urlprefix\url{https://link.aps.org/doi/10.1103/PhysRev.109.193}

\bibitem{PhysRev.109.1860.2}
E.~C.~G. Sudarshan, R.~E. Marshak,
  \href{https://link.aps.org/doi/10.1103/PhysRev.109.1860.2}{Chirality
  invariance and the universal fermi interaction}, Phys. Rev. 109 (1958)
  1860--1862.
\newblock \href {https://doi.org/10.1103/PhysRev.109.1860.2}
  {\path{doi:10.1103/PhysRev.109.1860.2}}.
\newline\urlprefix\url{https://link.aps.org/doi/10.1103/PhysRev.109.1860.2}

\bibitem{Mohapatra:1974gc}
R.~N. Mohapatra, J.~C. Pati, {A Natural Left-Right Symmetry}, Phys. Rev. D11
  (1975) 2558.
\newblock \href {https://doi.org/10.1103/PhysRevD.11.2558}
  {\path{doi:10.1103/PhysRevD.11.2558}}.

\bibitem{Mohapatra:1974hk}
R.~N. Mohapatra, J.~C. Pati, {Left-Right Gauge Symmetry and an Isoconjugate
  Model of CP Violation}, Phys. Rev. D11 (1975) 566--571.
\newblock \href {https://doi.org/10.1103/PhysRevD.11.566}
  {\path{doi:10.1103/PhysRevD.11.566}}.

\bibitem{Salam:1957st}
A.~Salam, {On parity conservation and neutrino mass}, Nuovo Cim. 5 (1957)
  299--301.
\newblock \href {https://doi.org/10.1007/BF02812841}
  {\path{doi:10.1007/BF02812841}}.

\bibitem{Kobzarev:1966qya}
I.~{\relax Yu}. Kobzarev, L.~B. Okun, I.~{\relax Ya}. Pomeranchuk, {On the
  possibility of experimental observation of mirror particles}, Sov. J. Nucl.
  Phys. 3~(6) (1966) 837--841, [Yad. Fiz.3,1154(1966)].

\bibitem{Foot:1991bp}
R.~Foot, H.~Lew, R.~R. Volkas, {A Model with fundamental improper space-time
  symmetries}, Phys. Lett. B272 (1991) 67--70.
\newblock \href {https://doi.org/10.1016/0370-2693(91)91013-L}
  {\path{doi:10.1016/0370-2693(91)91013-L}}.

\bibitem{Berezhiani:1995yi}
Z.~G. Berezhiani, R.~N. Mohapatra, {Reconciling present neutrino puzzles:
  Sterile neutrinos as mirror neutrinos}, Phys. Rev. D52 (1995) 6607--6611,
  [,279(1995)].
\newblock \href {http://arxiv.org/abs/hep-ph/9505385}
  {\path{arXiv:hep-ph/9505385}}, \href
  {https://doi.org/10.1103/PhysRevD.52.6607}
  {\path{doi:10.1103/PhysRevD.52.6607}}.

\bibitem{Blinnikov:1982eh}
S.~I. Blinnikov, M.~{\relax Yu}. Khlopov, {On Possible Effects of 'Mirror'
  Particles}, Sov. J. Nucl. Phys. 36 (1982) 472, [Yad. Fiz.36,809(1982)].

\bibitem{RevModPhys.90.045002}
G.~Bertone, D.~Hooper,
  \href{https://link.aps.org/doi/10.1103/RevModPhys.90.045002}{History of dark
  matter}, Rev. Mod. Phys. 90 (2018) 045002.
\newblock \href {https://doi.org/10.1103/RevModPhys.90.045002}
  {\path{doi:10.1103/RevModPhys.90.045002}}.
\newline\urlprefix\url{https://link.aps.org/doi/10.1103/RevModPhys.90.045002}

\bibitem{Arcadi:2017kky}
G.~Arcadi, M.~Dutra, P.~Ghosh, M.~Lindner, Y.~Mambrini, M.~Pierre, S.~Profumo,
  F.~S. Queiroz, {The waning of the WIMP? A review of models, searches, and
  constraints}, Eur. Phys. J. C78~(3) (2018) 203.
\newblock \href {http://arxiv.org/abs/1703.07364} {\path{arXiv:1703.07364}},
  \href {https://doi.org/10.1140/epjc/s10052-018-5662-y}
  {\path{doi:10.1140/epjc/s10052-018-5662-y}}.

\bibitem{Alexander:2016aln}
J.~Alexander, et~al., Dark sectors 2016 workshop: Community report, 2016.
\newblock \href {http://arxiv.org/abs/1608.08632} {\path{arXiv:1608.08632}}.

\bibitem{Battaglieri:2017aum}
M.~Battaglieri, et~al., Us cosmic visions: New ideas in dark matter 2017:
  Community report, 2017.
\newblock \href {http://arxiv.org/abs/1707.04591} {\path{arXiv:1707.04591}}.

\bibitem{Holdom:1985ag}
B.~Holdom, {Two U(1)'s and Epsilon Charge Shifts}, Phys. Lett. 166B (1986)
  196--198.
\newblock \href {https://doi.org/10.1016/0370-2693(86)91377-8}
  {\path{doi:10.1016/0370-2693(86)91377-8}}.

\bibitem{Glashow:1985ud}
S.~L. Glashow, {Positronium Versus the Mirror Universe}, Phys. Lett. 167B
  (1986) 35--36.
\newblock \href {https://doi.org/10.1016/0370-2693(86)90540-X}
  {\path{doi:10.1016/0370-2693(86)90540-X}}.

\bibitem{Berko1981787}
S.~Berko, K.~Canter, Encyclopedia of physics (1981) 787.

\bibitem{theoretical_ops}
G.~S. Adkins, R.~N. Fell, J.~Sapirstein,
  \href{https://link.aps.org/doi/10.1103/PhysRevLett.84.5086}{Order
  ${\ensuremath{\alpha}}^{2}$ corrections to the decay rate of
  orthopositronium}, Phys. Rev. Lett. 84 (2000) 5086--5089.
\newblock \href {https://doi.org/10.1103/PhysRevLett.84.5086}
  {\path{doi:10.1103/PhysRevLett.84.5086}}.
\newline\urlprefix\url{https://link.aps.org/doi/10.1103/PhysRevLett.84.5086}

\bibitem{rich_ops}
R.~S. Vallery, P.~W. Zitzewitz, D.~W. Gidley,
  \href{https://link.aps.org/doi/10.1103/PhysRevLett.90.203402}{Resolution of
  the orthopositronium-lifetime puzzle}, Phys. Rev. Lett. 90 (2003) 203402.
\newblock \href {https://doi.org/10.1103/PhysRevLett.90.203402}
  {\path{doi:10.1103/PhysRevLett.90.203402}}.
\newline\urlprefix\url{https://link.aps.org/doi/10.1103/PhysRevLett.90.203402}

\bibitem{Kataoka2009}
Y.~Kataoka, S.~Asai, T.~Kobayashi,
  \href{http://www.sciencedirect.com/science/article/pii/S0370269308014688}{First
  test of $\ensuremath{\mathcal{o}\left(\alpha^2\right)}$ correction of the
  orthopositronium decay rate}, Phys.~Let.~B 671~(2) (2009) 219 -- 223.
\newblock \href
  {https://doi.org/https://doi.org/10.1016/j.physletb.2008.12.008}
  {\path{doi:https://doi.org/10.1016/j.physletb.2008.12.008}}.
\newline\urlprefix\url{http://www.sciencedirect.com/science/article/pii/S0370269308014688}

\bibitem{Atoyan1989}
G.~S. "Atoyan, S.~N. Gninenko, V.~I. Razin, Y.~V. Ryabov,
  \href{"http://www.sciencedirect.com/science/article/pii/0370269389900592"}{"a
  search for photonless annihilation of orthopositronium "}, "Phys.~Lett.~B"
  "220"~("1--2") ("1989") "317 -- 320".
\newblock \href
  {https://doi.org/"http://dx.doi.org/10.1016/0370-2693(89)90059-2"}
  {\path{doi:"http://dx.doi.org/10.1016/0370-2693(89)90059-2"}}.
\newline\urlprefix\url{"http://www.sciencedirect.com/science/article/pii/0370269389900592"}

\bibitem{Mitsui1993}
T.~Mitsui, R.~Fujimoto, Y.~Ishisaki, Y.~Ueda, Y.~Yamazaki, S.~Asai, S.~Orito,
  \href{http://link.aps.org/doi/10.1103/PhysRevLett.70.2265}{Search for
  invisible decay of orthopositronium}, Phys.~Rev.~Lett. 70 (1993) 2265--2268,
  see Ref.~\cite{Gninenko1994} for a correction to the coupling constant limit
  due to collisions of oPs with matter.
\newblock \href {https://doi.org/10.1103/PhysRevLett.70.2265}
  {\path{doi:10.1103/PhysRevLett.70.2265}}.
\newline\urlprefix\url{http://link.aps.org/doi/10.1103/PhysRevLett.70.2265}

\bibitem{Badertscher2007}
A.~Badertscher, P.~Crivelli, W.~Fetscher, U.~Gendotti, S.~Gninenko, et~al., {An
  Improved Limit on Invisible Decays of Positronium}, Phys.~Rev.~D 75 (2007)
  032004.
\newblock \href {http://arxiv.org/abs/hep-ex/0609059}
  {\path{arXiv:hep-ex/0609059}}, \href
  {https://doi.org/10.1103/PhysRevD.75.032004}
  {\path{doi:10.1103/PhysRevD.75.032004}}.

\bibitem{Vigo2018}
C.~Vigo, L.~Gerchow, L.~Liszkay, A.~Rubbia, P.~Crivelli,
  \href{https://link.aps.org/doi/10.1103/PhysRevD.97.092008}{First search for
  invisible decays of orthopositronium confined in a vacuum cavity}, Phys. Rev.
  D 97 (2018) 092008.
\newblock \href {https://doi.org/10.1103/PhysRevD.97.092008}
  {\path{doi:10.1103/PhysRevD.97.092008}}.
\newline\urlprefix\url{https://link.aps.org/doi/10.1103/PhysRevD.97.092008}

\bibitem{Carlson:1987si}
E.~D. Carlson, S.~L. Glashow, {Nucleosynthesis Versus the Mirror Universe},
  Phys. Lett. B193 (1987) 168--170.
\newblock \href {https://doi.org/10.1016/0370-2693(87)91216-0}
  {\path{doi:10.1016/0370-2693(87)91216-0}}.

\bibitem{Berezhiani:2008gi}
Z.~Berezhiani, A.~Lepidi, {Cosmological bounds on the 'millicharges' of mirror
  particles}, Phys. Lett. B681 (2009) 276--281.
\newblock \href {http://arxiv.org/abs/0810.1317} {\path{arXiv:0810.1317}},
  \href {https://doi.org/10.1016/j.physletb.2009.10.023}
  {\path{doi:10.1016/j.physletb.2009.10.023}}.

\bibitem{Foot:2014mia}
R.~Foot, {Mirror dark matter: Cosmology, galaxy structure and direct
  detection}, Int. J. Mod. Phys. A29 (2014) 1430013.
\newblock \href {http://arxiv.org/abs/1401.3965} {\path{arXiv:1401.3965}},
  \href {https://doi.org/10.1142/S0217751X14300130}
  {\path{doi:10.1142/S0217751X14300130}}.

\bibitem{Berezhiani:2000gw}
Z.~Berezhiani, D.~Comelli, F.~L. Villante, {The Early mirror universe:
  Inflation, baryogenesis, nucleosynthesis and dark matter}, Phys. Lett. B503
  (2001) 362--375.
\newblock \href {http://arxiv.org/abs/hep-ph/0008105}
  {\path{arXiv:hep-ph/0008105}}, \href
  {https://doi.org/10.1016/S0370-2693(01)00217-9}
  {\path{doi:10.1016/S0370-2693(01)00217-9}}.

\bibitem{Foot:2015}
S.~V. R.~Foot, {Dissipative hidden sector dark matter}, Phys. Rev. D. 91 (2015)
  023512.
\newblock \href {http://arxiv.org/abs/1401.3965} {\path{arXiv:1401.3965}},
  \href {https://doi.org/10.1103/PhysRevD.91.023512}
  {\path{doi:10.1103/PhysRevD.91.023512}}.

\bibitem{Liszkay2008}
L.~Liszkay, C.~Corbel, P.~Perez, P.~Desgardin, M.-F. Barthe, T.~Ohdaira,
  R.~Suzuki, P.~Crivelli, U.~Gendotti, A.~Rubbia, M.~Etienne, A.~Walcarius,
  \href{http://aip.scitation.org/doi/abs/10.1063/1.2844888}{Positronium
  reemission yield from mesostructured silica films}, Applied Physics Letters
  92~(6) (2008) 063114.
\newblock \href
  {http://arxiv.org/abs/http://aip.scitation.org/doi/pdf/10.1063/1.2844888}
  {\path{arXiv:http://aip.scitation.org/doi/pdf/10.1063/1.2844888}}, \href
  {https://doi.org/10.1063/1.2844888} {\path{doi:10.1063/1.2844888}}.
\newline\urlprefix\url{http://aip.scitation.org/doi/abs/10.1063/1.2844888}

\bibitem{VigoPhD}
C.~Vigo, \href{https://doi.org/10.3929/ethz-b-000245118}{Search for invisible
  decay channels of positronium confined in a vacuum cavity}, Ph{D} {T}hesis,
  {ETH Zurich} (December 2017).
\newblock \href {https://doi.org/10.3929/ethz-b-000245118}
  {\path{doi:10.3929/ethz-b-000245118}}.
\newline\urlprefix\url{https://doi.org/10.3929/ethz-b-000245118}

\bibitem{Gidley1995}
D.~W. {Gidley}, D.~N. {McKinsey}, P.~W. {Zitzewitz}, {Fast positronium
  formation and dissociation at surfaces}, Journal of Applied Physics 78 (1995)
  1406--1410.
\newblock \href {https://doi.org/10.1063/1.360296}
  {\path{doi:10.1063/1.360296}}.

\bibitem{SIMION}
D.~A. Dahl,
  \href{http://www.sciencedirect.com/science/article/pii/S1387380600003055}{{SIMION}
  for the personal computer in reflection}, International Journal of Mass
  Spectrometry 200~(1) (2000) 3 -- 25, volume 200: The state of the field as we
  move into a new millenium.
\newblock \href
  {https://doi.org/http://dx.doi.org/10.1016/S1387-3806(00)00305-5}
  {\path{doi:http://dx.doi.org/10.1016/S1387-3806(00)00305-5}}.
\newline\urlprefix\url{http://www.sciencedirect.com/science/article/pii/S1387380600003055}

\bibitem{Geant4}
S.~Agostinelli, et~al., {GEANT4: A Simulation toolkit}, Nucl.~Instrum.~Meth.
  A506 (2003) 250--303.
\newblock \href {https://doi.org/10.1016/S0168-9002(03)01368-8}
  {\path{doi:10.1016/S0168-9002(03)01368-8}}.

\bibitem{Crivelli2010_2}
P.~Crivelli, U.~Gendotti, A.~Rubbia, L.~Liszkay, P.~Perez, C.~Corbel,
  {Measurement of the ortho-positronium confinement energy in mesoporous thin
  films}, Phys.~Rev. A81 (2010) 052703.
\newblock \href {http://arxiv.org/abs/1001.1969} {\path{arXiv:1001.1969}},
  \href {https://doi.org/10.1103/PhysRevA.81.052703}
  {\path{doi:10.1103/PhysRevA.81.052703}}.

\bibitem{Cassidy2010}
D.~B. Cassidy, P.~Crivelli, T.~H. Hisakado, L.~Liszkay, V.~E. Meligne,
  P.~Perez, H.~W.~K. Tom, A.~P. Mills,
  \href{http://link.aps.org/doi/10.1103/PhysRevA.81.012715}{Positronium cooling
  in porous silica measured via doppler spectroscopy}, Phys.~Rev.~A 81 (2010)
  012715.
\newblock \href {https://doi.org/10.1103/PhysRevA.81.012715}
  {\path{doi:10.1103/PhysRevA.81.012715}}.
\newline\urlprefix\url{http://link.aps.org/doi/10.1103/PhysRevA.81.012715}

\bibitem{NiRem}
N.~Oshima, R.~Suzuki, T.~Ohdaira, A.~Kinomura, T.~Narumi, A.~Uedono,
  M.~Fujinami, \href{https://link.aps.org/doi/10.1063/1.2919783}{Brightness
  enhancement method for a high-intensity positron beam produced by an electron
  accelerator}, J. Applied Physics 103 (2008) 094916.
\newblock \href {https://doi.org/10.1063/1.2919783}
  {\path{doi:10.1063/1.2919783}}.
\newline\urlprefix\url{https://link.aps.org/doi/10.1063/1.2919783}

\bibitem{Crivelli2010}
P.~Crivelli, A.~Belov, U.~Gendotti, S.~Gninenko, A.~Rubbia, {Positronium Portal
  into Hidden Sector: A new Experiment to Search for Mirror Dark Matter}, JINST
  5 (2010) P08001.
\newblock \href {http://arxiv.org/abs/1005.4802} {\path{arXiv:1005.4802}},
  \href {https://doi.org/10.1088/1748-0221/5/08/P08001}
  {\path{doi:10.1088/1748-0221/5/08/P08001}}.

\bibitem{Gninenko1994}
S.~N. Gninenko, {Limit on 'disappearance' of orthopositronium in vacuum},
  Phys.~Lett. B326 (1994) 317--319.
\newblock \href {https://doi.org/10.1016/0370-2693(94)91329-3}
  {\path{doi:10.1016/0370-2693(94)91329-3}}.

\end{thebibliography}
\end{document}